\documentclass{ringb99}
\usepackage{graphics}
\def\be{\begin{equation}}
\def\ee{\end{equation}}
\def\simlt{\ \raise -2.truept\hbox{\rlap{\hbox{$\sim$}}\raise5.truept   %MC
\hbox{$<$}\ }}                                                          %
\def\simgt{\ \raise -2.truept\hbox{\rlap{\hbox{$\sim$}}\raise5.truept   %
\hbox{$>$}\ }}   
\def\hmpc{{\rm\,h^{-1} Mpc}}
\begin{document}
\title{Cosmic Rays and Non-Thermal Emission in Galaxy Clusters}
\author{Sergio Colafrancesco\inst{1}}  
\institute{Osservatorio Astronomico di Roma, Via dell'Osservatorio 2,
I-00040, Monteporzio Italy}
\maketitle

\begin{abstract}
We discuss the origin of thermal and non-thermal phenomena in galaxy
clusters.
Specifically, we present some expectations for the non-thermal
emission (from radio to gamma ray wavelenghts) expected in a model in
which secondary electrons are produced via $pp$ collisions in the
intracluster medium.
We also discuss the possibility to test this cosmic ray model 
in the light of recent data on radio halos and hard X-ray
excesses found in nearby galaxy clusters.
\end{abstract}

\section{Introduction}
Clusters of galaxies emit thermal and non-thermal diffuse radiation in
the radio, EUV, X-rays, hard X-rays and possibly gamma rays.
In fact, 
the Coma cluster shows,  besides the well known thermal X-ray emission in the
$0.1 \div 10$ keV range, evidence of non-thermal diffuse
emission in the radio (see Feretti at this
Meeting) , EUV (see Lieu and Bowyer at this
Meeting)  and hard X-rays (see Fusco-Femiano at this Meeting).
In the gamma rays, there is only an upper limit $F(>100 MeV) = 4 \cdot
10^{-8} s^{-1} cm^{-2}$ obtained with EGRET (Sreekumar et al. 1996).

Several other clusters also show evidence of non-thermal emission and most of
them exhibit an extended radio halo.
All of these radio halo clusters have a strong merging
occurring in their central regions and do not show strong
cooling flows.
On the other hand, clusters with strong cooling flows are mostly
non-merging systems and do not show evidences for radio halos and/or
other non-thermal diffuse emission.
The cluster A2199 constitutes an exception: in fact, it has a strong
cooling flow in its center and a diffuse hard
X-ray and a soft X-ray/EUV excesses over the thermal model
(Kaastra et al. 1999). 

Most of the non-thermal phenomena occurring in clusters 
have been interpreted as due to synchrotron and
ICS emission from a population of relativistic cosmic ray electrons diffusing in the
intra cluster medium (ICM).
Cosmic ray models for the origin of non-thermal phenomena in clusters
belong basically to two categories: primary electron models (PEM; see, e.g., 
Jaffe 1977, Rephaeli 1979) in which relativistic electrons are
accelerated by discrete sources in the cluster, and possibly reaccelerated
by turbulence in the ICM (Jaffe 1977, Roland
1981, Schlickeiser et al. 1987); secondary electron models (SEM;
Dennison 1980, Vestrand 1982) in which cosmic ray electrons of galactic
origin are produced as secondary products of $pp$ collisions.
The main difficulty of the PEM is the short
propagation distance ($\sim 10 $ kpc) for the electrons.
This calls for continuous reacceleration processes in order to
explain diffuse non-thermal emission (radio and X-rays) out to Mpc
scales, as observed in Coma and A2199.
Other potential problems may come from the selective acceleration
efficiency required by non-hadronic models 
in order to avoid the production of
a large gamma ray emission coming from both the electron
bremsstrahlung and the $\pi^0 \rightarrow \gamma + \gamma$
electromagnetic decay.
The original SEM (Dennison 1980) has the potential problem of a
too steep injection spectrum (with index $\sim 2.5 \div 2.9$)
required to fit the Coma radio halo spectrum 
compared to the
standard CR spectral slopes ($\sim 2 \div 2.3$) (Chi et al. 1996).

In the following, we consider the available radio and X-ray data for clusters 
with known non-thermal
emission to discuss the possible constraints that can be set 
on the cosmic ray models for
the origin of such non-thermal phenomena.
Here we mainly discuss the
predictions from $pp$ collisions in the ICM.

\section{Theory}
We consider here a simple spherical cluster formed as a result of 
the collapse and virialization from an original top-hat perturbation 
originated at early times.
We assume that its final gas density 
distribution is given by $n(r)=n_c (1+x^2)^{-3 \beta/2}$ with 
$x\equiv r/r_c$,  where $n_c$ is the central density, $r_c$ is the cluster core radius and 
$\beta= \mu m_pv^2/kT$ (see Sarazin 1988 for a review).
If we assume virial and hydrostatic equilibrium of the gas in the cluster 
potential wells, then the IC gas temperature is  $T \propto M/R_v$, 
where $M$ is the total cluster mass and $R_v \propto r_c$ is its virial radius.
In the simple top-hat model, 
the cluster temperature and core radii are related by 
$r_c \propto T^{1/2} D^{3/2}(\Omega_0,z)$,
where the quantity 
$D(\Omega_0,z) = [\Delta(1,0)/ \Omega_0 \Delta(\Omega_0,z)]^{1/3} 
(1+z)^{-1}$ contains, in this model, 
all the cosmological dependence in terms of 
the evolution of the cluster density contrast,
$\Delta(\Omega_0,z) \equiv \rho_{cl}/\rho_b$ 
(here $\rho_{cl}$ is the internal density of the cluster at 
the collapse epoch and $\rho_b$ is the external background density;
see, e.g.,  Colafrancesco, Mazzotta \& Vittorio 1998 
for details).
The cluster X-ray luminosity is mainly provided by 
thermal bremsstrahlung emission and, in this framework, it reads
$L_X \propto n_c^2 r_c^3 T^{1/2}$.

\subsection{Non-thermal emission}
Cosmic rays (CR) produced within a cluster can yield gamma-rays
through the production and the subsequent decay of neutral pions:
$
p+p\to \pi^0+X$ with $\pi^0\to \gamma+\gamma.
$
In the same interactions, charged pions are also produced, which 
determine neutrino and $e^{\pm}$ emission through 
$
p+p\to\pi^{\pm}+X$, $\pi^{\pm}\to \mu^{\pm} 
\nu_{\mu}(\bar{\nu}_{\mu})$, $\mu^{\pm}\to e^{\pm} + 
\bar{\nu}_{\mu}(\nu_{\mu})
+ \nu_e (\bar{\nu}_e)$.
These secondary products occur at the same place ({\em e.g., in situ}) 
of $pp$ collisions.

Various sources of CRs are effective in clusters. Normal galaxies provide at most
CR production rates $\sim 10^{43}$ erg/s  and active galaxies can yield $\sim 10^{44}$
erg/s.
Also ICM shocks, produced by accretion and/or merging in clusters, 
are effective in accelerating CR and
can yield $\simgt$ a few $10^{44}$ erg/s (see CB98 for details). 
These CR production rates are averaged over a Hubble time.

The CR diffusion and confinement in the IC medium are
{\it crucial} mechanisms  to maximize the efficiency of $pp$ 
collisions and are strictly related to the value and 
configuration of the magnetic field $B$ in clusters.
For a Kolmogorov spectrum one obtains (CB98):
\be
D(E_p)
%=\frac{1}{3} c d_0^{2/3} (eB)^{-1/3} E^{1/3} 
\approx 
2.3\cdot 10^{26}~E_{GeV}^{1/3}~B_{\mu G}^{-1/3}~cm^2/s ~.
\ee
For the range of energies we are interested in,
CRs propagate {\em diffusively} according to the transport
equation:
\be
D(E_p) \nabla^2 n_p(E_p,r) - \frac{\partial}{\partial E_p}
\left[ b(E_p) n_p(E_p,r)\right] = Q(E_p)\delta (\vec {r})
\label{eq:transport}
\ee
where
$n_p(E_p,r)$ is the density of CRs with energy $E_p$,
$Q_p(E_p) = Q_0 E_p^{-\gamma}$ is the spectrum of the CR source in the cluster
and $b(E_p)$ is the rate of CR energy losses.
For average IC gas densities $\bar{n}\sim 3 \cdot 10^{-4}~cm^{-3} ~h^2$ (as in Coma),
$x_{cl} \propto {M_{gas} \over R_v^2} \sim 6~g/cm^2 \ll 
x_{nucl}=(m_p/\sigma_{pp})\approx 50 \div 100~g/cm^2$ and
CR energy losses are $\approx 0$.
In the case of a point-like CR source
the solution of eq.(2) is 
\be
n_p(E_p,r)=\frac{Q_p(E_p)}{D(E_p)}\frac{1}{2\pi^{3/2} r}
%\int_{\sqrt{\frac{r^2}{6D(E_p)t_0}}}^{\infty} dy e^{-y^2} ~,
\int_{r/r_{max(E_p)}}^{\infty} dy e^{-y^2}  ~.
\label{eq:sol}
\ee
The maximum length over which CR can diffuse in a Hubble time, $t_0$, is: 
\be
r_{max} \approx 0.5 \hmpc ~~\bigg({ t_0 \over 2 \cdot 10^{10} yr} \bigg)^{1/2}
\cdot \bigg( { D \over 10^{29} cm^2 ~s^{-1}} \bigg)^{1/2}
\ee
and the CR diffusion time at energy $E$ and distance $r$ is:
\be
\tau_D=\frac{r^2}{4 D(E_p)} 
	\approx 2.2\times 10^{5} Gyr ~r_{Mpc}^2 E_{GeV}^{-1/3} ~.
\ee
Thus, confinement on  scales $R_{cl}\approx 2 \hmpc$ happens for
$E_p \simlt 4.2\times 10^{5} ~ GeV$ and
confinement in the core, $r_c \approx 0.2 \hmpc$ 
is effective for $E_p \simlt 8.8\times 10^{3} ~ GeV$.

In the extreme case of a homogeneous injection of CR in the
ICM, the equilibrium CR distribution is given by
\begin{equation}
n_p(E_p)=K\frac{\epsilon_{tot}}{V}p_p^{-\gamma} ~,
\label{eq:sol_homo}
\end{equation}
where the constant $K$ is obtained requiring that
$
K\int_0^{E_p^{max}} dT_p T_p p_p^{-\gamma}=\epsilon_{tot}~,
$
and $\epsilon_{tot}$ is the total CR energy injected in the cluster
volume $V$.

Once CR protons are confined in the ICM, they can produce secondaries through $pp$
collisions with the thermal protons in the ionized, metal rich ICM.
The number of i-secondaries ($i=\gamma, \nu$) per unit time 
and unit volume, at energy $E_p$ 
is $q_i(E_p,r) = K ~Y_i~ n_p(E_p,r)$ (BBP, CB98).
In the diffusion regime,
the total number of i-secondaries reads:
\be
Q_i(E_p) \approx K ~\bigg[ Y_i Q_p(E_p) \bigg] ~\tau_D ~,
\ee
where
$K = n \sigma_{pp} c$ is the rate of $pp$ collisions,
$\sigma_{pp} \approx 3.2 \times 10^{-26} cm^{-2}$ is the $pp$ cross section,
and $Y_i$ are the yields for i-secondary production.  

The emissivity in gamma rays at distance $r$ and
energy $E_\gamma$ is given by
$$
j_\gamma^{\pi^0}(E_\gamma,r)=2 n(r) c
\int_{E_\pi^{min}(E_\gamma)}^{E_p^{max}}
dE_\pi 
$$
\be
\int_{E_{th}(E_\pi)}^{E_p^{max}}
dE_p F_{\pi^0}(E_\pi,E_p)\frac{n_p(E_p,r)}{(E_\pi^2+m_\pi^2)^{1/2}},
\label{eq:gamma1}
\ee
where $E_\pi^{min}(E_\gamma)=E_\gamma+m_{\pi^0}^2/(4E_\gamma)$
(see BC99 for further details).

The production spectrum of secondary electrons, $q_e(E_e,r)$, 
can be derived from the known $\pi^{\pm}$, $\mu^{\pm}$ spectra
(see BC99 for details).
The equilibrium electron spectrum is obtained from the transport equation:
\be
D(E_e) \nabla^2 n_e - \frac{\partial}{\partial E_e}
\left[ b(E_e) n_e\right] = q(E_e)\delta (\vec {r}) ~,
\label{eq:transport}
\ee
where $b_e(E_e)$ is the rate of energy losses at energy $E_e$.
%The equilibrium spectrum for $e^{\pm}$ reads:
%\be
%n_e(E_e,r) = \frac{q_e(E_e,r)}{(\gamma+\eta-1)}\frac{E_e}{b_e(E_e)}
%\ee
These secondary electrons produce 
synchrotron radio emission in the magnetized ICM
\be
j(\nu) = \left\{ n_e \left(\frac{dE_e}{dt}\right)_{syn}~  
\frac{dE_e}{d\nu} \right\} ~,
%
%\propto ~ n(r) c \sigma_0 Y_{\gamma}
%\cdot \frac{Q_0}{b_0(B_{\mu})} B_{\mu}^{\frac{\gamma+\eta+2}{2}}~
%\nu^{-\frac{\gamma+\eta}{2}}
\ee
and X-ray emissivity through Inverse Compton Scattering
(ICS) off the microwave background photons
\be
\phi_X(E_X) = \left\{ n_e \left(\frac{dE_e}{dt}\right)_{ICS}~
\frac{dE_e}{dE_X}  \right\}~
%\propto ~
%n(r) c \sigma_0 Y_{\gamma}
%\cdot \frac{Q_0}{b_0(B_{\mu})}~
%E_X^{-\frac{\gamma+\eta}{2}}
\ee
(see Longair 1993 for details).
The calculation of the radio emissivity is
performed here in the simplified assumption that electrons with energy $E_e$
radiate at a fixed frequency given by:
\begin{equation}
\nu \approx 3.7\cdot 10^6 B_{\mu} E_e^2(GeV) Hz.
\label{eq:freq}
\end{equation}
For ICS also
we adopt the approximation that electrons radiate at a single energy,
given by
\begin{equation}
E_X=2.7~keV ~E^2_e(GeV).
\label{eq:ex}
\end{equation}
These approximations introduce negligible errors in the final result and
have the advantage of making it of immediate physical interpretation.

\section{Observational constraints}
The most natural prediction of the SEM  is a substantial
diffuse emission of gamma rays  extending at least up to $\simgt 10$ GeV (see Fig.1).
On the other hand, the PEM predicts a substantially lower gamma-ray flux from clusters
due to relativistic bremsstrahlung from high energy electrons and moreover with a
steeper spectrum which is strongly suppressed at energies $\simgt 1$ GeV.

A definite prediction of the SEM is the existence of a correlation between the gamma
ray flux, $F_{\gamma}$, and the thermal X-ray flux, $F_X$, of clusters
\be
F_{\gamma} / F_X \propto 
Y_{\gamma} \sigma_{pp} c
\left( \frac{1}{n_c r_c} \right) \frac{1}{(kT)^{1/2}} 
\left\{ \frac{Q_p(E)}{D_{CR}(E)} \right\}
\ee
(see CB98 for details) which arises because of the dependence of the rate of $pp$
collisions from the density of the gas (see eq.7).
The gamma ray flux predicted for Coma from eq.(14), $F(>100 ~MeV) = 8.5 \cdot 10^{-9}
~photons ~s^{-1} ~cm^{-2}$, and those of many nearby clusters which 
are also in the range 
$5 \cdot 10^{-10} \div 10^{-8} ~photons ~s^{-1} ~cm^{-2}$ are unobservable
with the present  gamma ray telescopes.
The cluster gamma ray spectrum of Coma predicted in the SEM 
shows also a peculiar feature (bump) at $\sim
70$ MeV which is directly correlated to the $\pi^0 \rightarrow \gamma + \gamma$ decay
(the $\pi^0$ mass is $\approx 140$ MeV).
\begin{figure}
 \resizebox{\hsize}{!}{\includegraphics{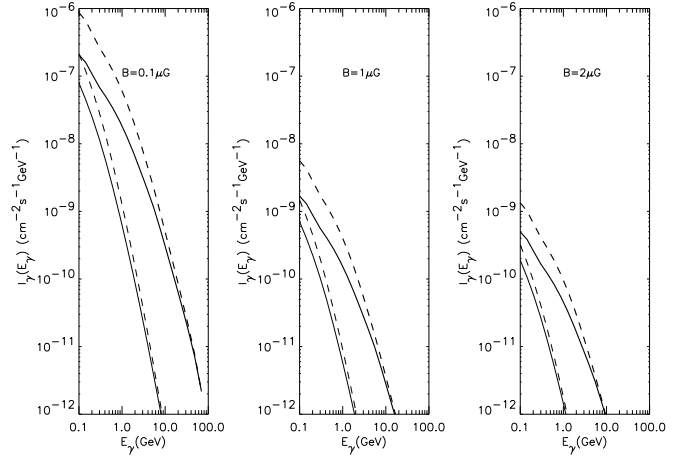}}
\caption[]{The three panels refer
to the values of $B$ as indicated. The thick lines represent
the contribution of neutral pion decay, while the thin lines represent
the bremsstrahlung contribution of secondary electrons. Solid and
dashed curves are for $\gamma=2.1$ and $\gamma=2.4$, respectively.
}
\end{figure}
The next generation gamma ray telescope (GLAST, VERITAS) have enough sensitivity and
spatial resolution to disentengle between the role of the SEM and of the PEM in galaxy
clusters.
These facilities will be however available in the next decade.
Meanwhile, the available data in the
radio and hard X-ray energy ranges can be used
to put constraints on the cosmic ray models for the
origin of non-thermal processes in clusters.

\subsection{Constraints from the Coma cluster}
The SEM is able to reproduce the integrated spectrum of the Coma radio halo
up to $1.4$ GHz for different values of the $B$ field (see Fig.2), 
contrary to the claims by Chi et al. (1996).
The high-$\nu$ data could be also reproduced using a more 
detailed diffusion picture for CR in the ICM.
\begin{figure}
 \resizebox{\hsize}{!}{\includegraphics{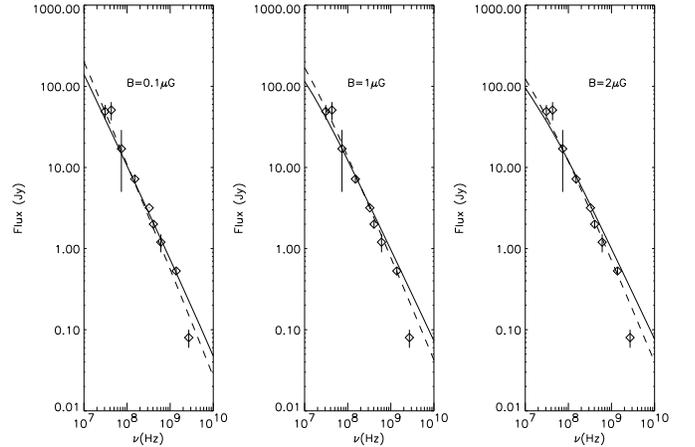}}
\caption[]{Spectrum of the diffuse radio emission from Coma for
values of the average intracluster magnetic field as indicated.
%$B=0.1~\mu G$ (left panel), $B=1~\mu G$ (central panel) and
%$B=2~\mu G$ (right panel).
A King density profile has been used with
$\beta=0.75, n_c=2.89 \cdot 10^{-3} ~cm^{-3}$ and $r_c=0.42$ Mpc
(we use  $h=1/2$).
For each panel the cases $\gamma=2.1$ (continuous curves) and
$\gamma=2.4$ (dashed curves) are shown.}
%Data from Giovannini et al. (1993).}
\end{figure}
The observed Hard X-Ray Excess (HXE) in Coma (Fusco-Femiano et al. 1999)
can be reproduced in the SEM only with low values of $B$ (see BC99).
%
%\begin{figure}
% \resizebox{\hsize}{!}{\includegraphics{/user2/cola/bc99_1_f2new.ps}}
%\caption[]{Spectrum  of the diffuse X-ray emission from Coma.
%The three panels refer to the same values of the IC magnetic field as
%in the previous Figure.
%The shaded area shows the best fit to the HEAO1-A4 and GINGA
%thermal emission data (open triangles) at $T=8.21 \pm 0.20$ keV.
%The OSSE upper limits  are indicated by the open circles.
%The SAX data  are indicated by filled squares.
%Arrows and labels show, for each panel, the energy ranges in which the
%three different data sets are located.
%Predictions of the SEM for $\gamma=2.1$ (continuous curves) and
%$\gamma=2.4$ (dashed curves) are shown in each panel}
%\end{figure}
%
The Coma HXE, if confirmed, puts a strong constraint to the SEM.
In fact, if the Coma HXE is mostly produced by ICS, the SEM should be ruled out because
the gamma ray flux predicted for Coma in the case $B \approx 0.1 \mu G$ (the case which
fits both the Coma radio halo spectrum and its HXE) exceeds the EGRET upper limit 
by a factor $\sim 2 \div 7$ (see Table 1).
\begin{center}
\begin{table}
\caption{
\label{table1}}
\vskip 0.1truecm
\begin{tabular}{| c | c | c | c | c |}
\hline
\hline
$B_\mu$ & $\gamma$ & $\frac{L_p}{10^{44}erg s^{-1}}$ &
$\frac{F_\gamma(E_\gamma\geq 100MeV)}{F_\gamma^{EGRET}
(E_\gamma\geq 100MeV)}$ & $F_{\gamma}^{brem}/F_{\gamma}^{\pi^0}$
\\ \hline
$0.1$ & $2.1$ & $50$ & $1.93$ & 0.13\\
$0.1$ & $2.4$ & $180$ & $7.15$ & 0.10 \\
$1$ & $2.1$ & $0.35$ & $1.8\cdot 10^{-2}$ & 0.14 \\
$1$ & $2.4$ & $1$ & $4.5\cdot 10^{-2}$ & 0.11\\
$2$ & $2.1$ & $0.1$ & $5.3\cdot 10^{-3}$  & 0.12\\
$2$ & $2.4$ & $0.23$ & $1.1\cdot 10^{-2}$ & 0.095\\
\hline
\hline
\end{tabular}
\end{table}
\end{center}
However, if
the magnetic field in Coma decreases with the distance - as indicated by recent
simulations (see Friaca \& Goncalves 1999) -
from $\sim$ a few $\mu$G
at $\sim 1$ Mpc, as probed by the radio halo, to $\sim 0.1 \mu$G at $\sim 3$ Mpc, as
probed by the SAX observation, 
then the SEM could be consistent with radio,
HXE and gamma ray data on Coma.
Nonetheless, the HXE observed by SAX can be interpreted more simply by bremsstrahlung
emission of a population of high energy electrons which are stochastically
accelerated in the ICM and populate the supra-thermal
tail of a Maxwellian distribution (see Dolgiev 1999, Ensslin 1999 in these Proceedings).
If the HXE in Coma can be explained in this last way, then the SEM would be entirely consistent
with the radio, X-ray and gamma-ray data.

\subsection{The case of A2199}
A soft excess in EUVE-DS, ROSAT-PSPC, SAX-LECS together with a HXE in the
SAX-PDS data have been found recently in A2199 (Kaastra et al. 1999).
The HXE in A2199 has been also detected in the SAX-MECS, which allowed 
to map the increase of the HXE 
from the inner regions out to $\sim 24$ arcmin away from the
cluster center (see Fig.3). 
While the HXE in the inner $3$ arcmin of A2199 
could be due to the steep spectrum, central radio galaxy, 
the HXE detected in the outer parts of A2199, requires unavoidably an
explanation in terms of a diffuse non-thermal emission.

At energies $\sim 30$ GeV, the maximum distance that protons can diffuse
away from the central CR source is: 
\be
r_{max}(E) \approx
0.8 ~ Mpc ~ B_{\mu}^{-1/6} \bigg({l_c \over 20
kpc}\bigg)^{1/3} \bigg({t_0 \over 15 Gyr}\bigg)^{1/2} ~.
\ee
Protons can reach distances $\sim 1.2$ Mpc, at which the HXE is
observed, if the average value of the magnetic field is $B_{\mu}
\simlt 0.08$. This low value of $B$ allows the diffusion of
CR protons 
in the cluster out to large distances 
without affecting the IC emission induced by secondary electrons.

In the case of a central CR source, 
the IC emissivity integrated along the line of sight,
$l$, and
evaluated at the projected distance $\theta \approx r/d_A(\Omega_0,z)$
[here $d_A(\Omega_0,z)$ is the angular diameter distance], is
\be
\phi(E_X,\theta) \propto \int dl r^{-1} n(r) \int_{r/r_{max}}^{\infty}
dy ~exp(-y^2) ~.
\ee
\begin{figure}
 \resizebox{\hsize}{!}{\includegraphics{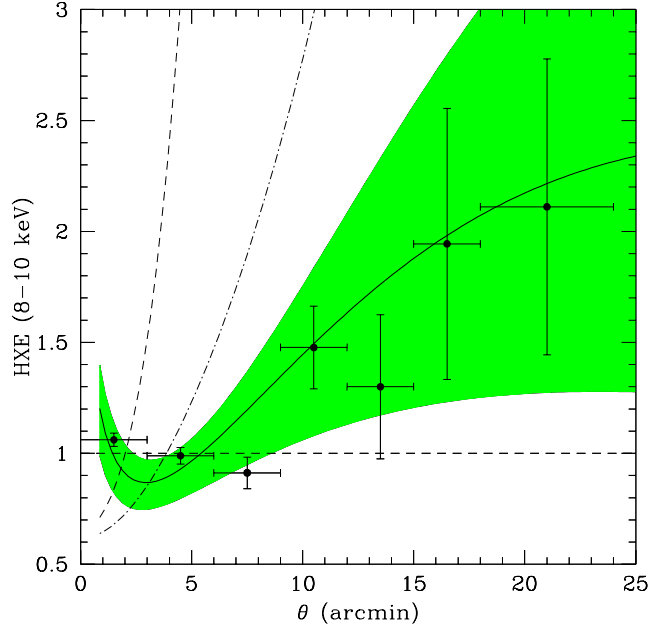}}
\caption[]{The HXE in the $8-10$ keV energy band in the 
different models discussed in the text.
Solid curve shows the prediction of the SEM in
the case of a central CR source.
The shaded area represent the theoretical uncertainty region whene the
observed uncertainties in $\theta_c$ and $\beta$ for
A2199 are taken into account.
Data are from Kaastra et al. (1999).
The dashed curve shows the behaviour $HEX \propto n^{-2}(\theta)$
and the dot-dashed curve the behaviour $HEX \propto n^{-1}(\theta)$.
}
\end{figure}
The projected bremsstrahlung emissivity is
$\varepsilon(E_X,\theta) \propto \int dl~ n^2(r)$ if $T=$ const.
The ratio HXE $= \phi(\theta)/\varepsilon(\theta)$,
evaluated numerically in the range $8 - 10$ keV, 
is shown in Fig.3 for the best fit values $\theta_c=3.7$
arcmin and $\beta=0.78$.
Remarkably, the SEM is able to reproduce
the observed spatial behaviour of the HXE at both small and
large radii.
The small radius excess is due to the presence of the central CR source
whose effect is described by a peaked distribution of CR protons,
$n_p(E_p,r) = Q_p(E_p)/[4 \pi D(E_p) r]$ (the solution of the proton diffusion
equation for $r \ll r_{max}$), which reflects in an analogous
behaviour of the secondary electron IC emissivity.
The large radius excess is mainly provided by the flatter
spatial dependence of the IC emissivity, 
$\phi \sim Q_p(E_p)n(r)/D(E_p) r$, compared to the much rapid decrease
of the thermal bremsstrahlung emissivity,
$\varepsilon \sim n^2(r) T^{1/2}$.

In the PEM, 
the relativistic electrons produced by the central radiogalaxy suffer
strong energy losses which prevent them from diffusing out to
distances larger than  a few dozens kpcs.
A possible strong IC emission from the radio galaxy would then explain
the small radius excess of A2199 but not the large radius excess.
If a reacceleration process is effective in A2199 out to $\sim 1$ Mpc
from its center, then the predicted HXE increases as
\be
HXE = {\phi^{PEM} \over \varepsilon}  \propto {N_e \over n^2(\theta)
T^{1/2}} ~.
\ee
If the relativistic electron density, $N_e$, is uniformly accelerated in the
ICM, then the HXE $\propto 1/n^2(\theta)$, which
increases much faster than the observed HXE.
If, however, the relativistic electrons are accelerated at large
distance from the center of A2199 and soon reach equipartition with the
thermal gas, then 
%$N_e \sim 1.5(x-1)(kT/E_{min}) n(r)$, and the
the predicted HXE increases as 
\be
HXE = {\phi^{PEM} \over \varepsilon}  \propto {1 \over n(\theta)
T^{1/2}} ~,
\ee
which is still too steep than the observed HXE (see Fig.3).
If the  HXE in A2199
is due to bremsstrahlung emission from a population of high
energy electrons, 
%different regimes are possible in this case.
%In the case of high energy, non-relativistic electrons with a
%power-law distribution of energies,
%$N_e=N_0E^{-x}$, the emissivity is
%$
%\varepsilon_{rel} \propto Z^2 n(r) N_e {E \over (3/2 -x)}
%$
%(see Longair 1994)
%which yields HXE $\propto [N_e/n(r)] (E/T^{1/2})$.
%In the case of bremsstrahlung emission from a population of
%relativistic electrons with a power-law  spectrum, 
%$N_e=N_0E^{-x}$, the emissivity is 
%$
%\varepsilon_{rel} \propto Z(Z+1.3) n(r) N_e {E^2 \over (2 -x)}
%$
%(see Longair 1994).
%This yields HXE $\propto {Z+1.3 \over Z} 
%[N_e/n(r)] (E^2/T^{1/2})$.
the spatial behaviours $HXE \sim 1/n(\theta)$ and $HXE =$ Const obtain
when the electron spectrum, $N_e$, is uniformly distributed in the cluster or when
equipartition of relativistic electrons and thermal gas is assumed,
respectively.

The previous results indicate that ICS of secondary electrons 
can reproduce naturally the spatial dependence of the HXE
observed in A2199.
The  solution in eq.(3) is able to reproduce both the small radius and the large
radius excesses in A2199.
The CR power needed to reproduce also the amplitude of the observed 
HXE in A2199 is of order of a few $10^{45} erg/s$, a value slightly in
excess over the average production rate estimated for active galaxies in
clusters (BC98).
An important point to notice is that the SEM  naturally
predicts that the HXE should be accompanied by an extended radio halo
due to synchrotron emission of the same electrons in the intracluster magnetic
field.
The power, $j(\nu)$, of such radio halo is extremely
low due to the strong dependence, $j(\nu) \sim B^{1+\alpha_r}$. Using
$B \simlt 0.08 ~\mu G$, as required to diffuse protons out
to $\sim 1.2$ Mpc from the cluster center and
assuming $L_p \approx 10^{44}$ erg/s, we expect a luminosity,
$J_{1.4} \sim (1.9 \div 4.0) \cdot 10^{20} $ W/Hz 
%($h=1/2$), 
for the A2199 radio halo, which gives a vary faint brightness, 
below the current observational sensitivities.
Simple models of primary electron reacceleration have difficulties in
reproducing the observed HXE because they predict a too flat profile of
IC emission compared to the thermal bremsstrahlung one.
These models require a inhomogeneous reacceleration mechanism which 
has to be more
efficient at large distances from the A2199 center and a condition of
almost complete equipartition with the thermal gas.
Bremsstrahlung emission from a non-thermal or supra-thermal 
population of electrons could explain the HXE if a
stochastic acceleration of the high-energy electrons
results to be more efficient at large distance from the cluster center.

\section{The radio luminosity - temperature correlation}
The available data on the radio halo clusters 
(see Giovannini \& Feretti 1996, Feretti 1999, Liang 1999, these Proceedings) 
show a tight correlation between their radio power
at $1.4$ GHz, $J_{1.4}$, and their ICM temperature, $T$ (see Fig.4).
Different temperature data (we consider the D93
temperatures with the associated $90 \%$ c.l. uncertainties 
and for the AE99 and PXF99 data, we quote  the $2$ sigma 
uncertainties on $T$) are quite consistent among them and 
do not introduce a systematic large 
scatter in the $J_{1.4}-T$ relation.
Fitting these data with a simple power law, $T = A J_{1.4}^b$, 
we obtain the following best fit parameters 
(we give in parentheses the $90 \%$ c.l. range):
$A=6.94$ (range $6.13 \div 7.84$) and $b=0.16$ (range $0.08 \div
0.24$) 
with a $\chi^2_{red}=0.87$, for the AE99 temperature data.
The results of the fit do not change very much if we use the D93 data
[$A=6.76$ (range $5.41 \div 7.93$) and $b=0.18$ (range $0.14 \div
0.24$) with a $\chi^2_{red}=2.48$] or the PXF99 data [$A=7.12$ (range
$6.40-8.02$) and $b=0.16$ (range $0.08 \div 0.22$) with a
$\chi^2_{red}=0.80$].

This correlation can be used to put constraints on models for the radio halo origin.
In the case of a single CR source acting 
in the central regions of the cluster,
the integrated radio halo spectrum predicted in the SEM is 
\be
J^{SEM}(\nu) \propto L_p B^{1+\alpha_r} n_c r_c^2 \nu^{-\alpha_r} ~,
\label{eq:eq_jsem}
\ee
where $L_p$ is the cluster production rate of CR.

%The monochromatic power predicted in the PEM depends, instead, 
%only on the spectrum, $n_e(E_e)$, of 
%the primary CR electrons accelerated in the ICM and on the magnetic
%field and is
%\be
%J^{PEM}(\nu) \propto N_e \nu^{-\alpha_r} ~,
%\label{eq:eq_jpem}
%\ee
%where $N_e \propto \int dE_e n_e(E_e)$ is the total number of electrons per 
%unit volume.

Substituting $n_c \propto L_x^{1/2} r_c^{-3/2} T^{-1/4}$ (as  derived from thermal 
bremsstrahlung emission) in eq.(19) one obtains:
\be
J(\nu) \propto {L_X^{1/2} r_c^{1/2} \over T^{1/4} } \cdot L_p
B^{1+\alpha_r} \nu^{-\alpha_r}
\label{eq:jgen}
\ee
which simplifies to 
$J(\nu) \propto L_p L_X^{1/2}B^{1+\alpha_r} \nu^{-\alpha_r}$ 
for a spherical, virialized cluster.
Using the observed  correlation $L_X \propto T^{\alpha}$ 
with $\alpha \sim 2.88 \div 3.45$ (we use here the best fit slopes given by 
AE99 and  D93, respectively) one finally finds, in the framework of the
SEM, the relation
\be
J(\nu) \propto D^{3/4} L_p T^{\alpha/2} B^{1+\alpha_r} \nu^{-\alpha_r} ~.
\ee
The predictions of the simple top-hat model in which 
$B$ and $L_p$ do not depend on $T$ cannot fit the data.
On the other hand, the effects of merging play a relevant role in the halo formation.
First, the merging clusters found in the AE99 sample here considered
show evidence of a core radius sensibly larger than that 
of normal clusters.
In particular, 
a relation, $r_c \sim T^{1/2+\epsilon}$ with 
$\epsilon \approx 0.5 \div 1$ is indicated by the available data 
and expected on simple theoretical grounds (see Colafrancesco 1999).
The merging correction to $r_c$, by itself,
increases sensibly the quality of the fit in the SEM but it does not 
reproduce the observed $J_{1.4}-T$ correlation.

The data require a strong $T$ dependence of $L_p$ and/or $B$.
A possibility to recover the SEM is to assume that 
the ICM shocks occurring in radio halo clusters are pervasive and 
occupy all the cluster volume with a 
continuous and efficient set of reacceleration events.
%In this case, the PEM specific radio power should be 
%integrated over the cluster volume providing a relation 
%$J_{1.4} \sim T^{3/2}$ which gives a very marginal
%%($\chi^2_{red}\approx 11.4$) 
%fit to the data.
In the case of a very extended and continuous CR 
source, the SEM would predict a relation $J_{1.4} \sim
T^{(\alpha+1+3\epsilon)/2}$ which is sensibly steeper 
%($J_{1.4} \sim T^3$ for $\alpha=3.45$ and $\epsilon=0.5$, and 
($J_{1.4} \sim T^{3.8}$ for $\epsilon=1$)
than that for a single CR source.
In this framework, moreover, 
merging can also provide a non-universal CR production rate which
increases with increasing cluster temperature (or mass).
As a simple example, let us consider that a fraction $f$ of the 
kinetic energy involved in the merging of two subclumps, $E_{merg} \propto
M_1v_1^2 + M_2v_2^2$ (with $M_1+M_2=M$, for simplicity),
is transferred to the accelerated particles so that, $L_p \sim f
E_{merg}/\tau_{acc} \propto M \sim T^{3/2}$
(we consider here for the sake of illustration the case of 
a constant acceleration time scale, $\tau_{acc} =$
const, and uncorrelated relative velocities, $v_1$ and $v_2$).
%A similar argument applies also in the PEM to give 
%$N_e \propto T^{3/2}$.
Adding also this ingredient, consistently with the merging picture,
the SEM predicts a relation $J_{1.4} \sim T^{5.2}$
(for $\alpha=3.45$ and $\epsilon =1$) which fits well 
%($\chi^2_{red} \approx 2.1$) 
the observed $J_{1.4}-T$ correlation (see Fig.4).
Also the PEM prediction in this case is  
$J_{1.4} \sim T^{3}$ but is still not sufficient 
%($\chi^2_{red} \approx 8.5$) 
to fit the data.
However, assuming also an overall equipartition (both with the thermal
gas and with the magnetic field) a steeper relation,
$J_{1.4} \sim T^{3.4 \div 4.2}$, obtains in the PEM (Colafrancesco 1999).
\begin{figure}
 \resizebox{\hsize}{!}{\includegraphics{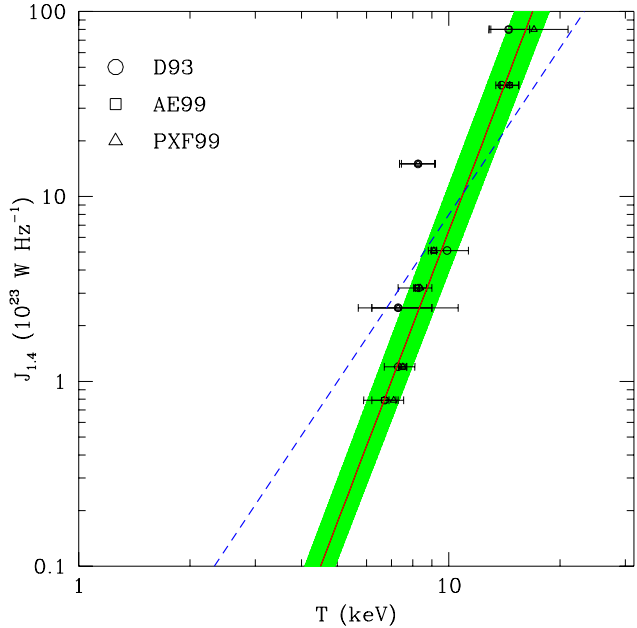}}
\caption[]{The best fit relation (solid line) for radio halo clusters including 
the merging correction $r_c \sim T^{1/2+\epsilon}$ 
in the case of a pervasive shock with the condition 
$L_p \sim T^{3/2}$.
Monochromatic radio powers are taken from Giovannini \& Feretti 
(1996-1999) and Liang (1999)
and temperatures (with $90 \%$ uncertainties) from 
PXF99 (open triangles), AE99 (open squares) and
D93 (open circles).
The shaded area shows the $95.4 \%$ uncertainties on the best fit
parameters.
The case $N_e \sim T^{3/2}$ and
overall equipartition between magnetic field and IC gas energy
densities in the PEM is shown by the dashed line.
Here, the radio luminosities $J_{1.4}$ are given for
$H_0=100 ~km ~s^{-1} ~ Mpc^{-1}$.}
\end{figure}
\section{Conclusions}
The available data on the radio and hard X-ray emission in galaxy clusters 
provide a test-bed for cosmic ray models for the origin of such
non-thermal processes.
The SEM yields a viable alternative to explain the overall characteristics of the
radio halo spectra and the consistency of the gamma-ray emission with the EGRET limit
of Coma, provided that the bulk of the HXE observed in this cluster is due to
stochastic acceleration of suprathermal electrons.
Moreover, the SEM provides also a viable explanation of the spatial distribution of
the HXE observed in A2199.
Simple primary electron 
models have difficulties in reproducing the spatial extent of the HXE
in A2199, unless an efficient spatially inhomogeneous acceleration is effective in
the ICM of this cluster.
Both the SEM and the PEM moreover require a non-universal CR production rate to
explain the observed $J_{1.4}-T$ relation.

The extensive search for radio halos and HXE in low and high-$T$ 
clusters, together with the
future observations of the gamma-ray emission from nearby clusters 
which will be available with the future high energy experiments 
(GLAST, VERITAS, MAGIC, STACEE), will definitely break the model 
degeneracy here discussed and probe the origin of non-thermal phenomena in 
galaxy clusters.

%\begin{acknowledgements}
%We gratefully acknowledge the permission by the Springer Verlag
%to use their A\&A \LaTeX{} document class macro.
%\end{acknowledgements}


\begin{thebibliography}{}
\bibitem{ae99} Arnaud, M. \& Evrard, G. 1999, A\&A, in press (AE99)
\bibitem{bbp} Berezinsky,V., Blasi,P. \& Ptuskin,V.1997, ApJ, 487, 529 (BBP)
\bibitem{bc99} Blasi, P. \& Colafrancesco, S. 1999, APh., in press (BC99)
\bibitem{chi96} Chi, X. et al. 1996, Phys. Rev. Lett., 77, 1436 
\bibitem{cola99} Colafrancesco, S. 1999, ApJ, submitted
\bibitem{cmv98} Colafrancesco, S. et al. 1998, ApJ, 488, 566
\bibitem{cb98} Colafrancesco, S. \& Blasi, P. 1998, APh., 9, 227 (CB98)
\bibitem{d93}  David, L.P. et al. 1993, ApJ, 412, 479 (D93)
\bibitem{den80} Dennison, B. 1980, ApJ, 239, L93
\bibitem{fusco99} Fusco-Femiano, R., et al. 1999, ApJ, in press
\bibitem{fg99} Friaca, F. \& Goncalves, G. 1999, preprint astro-ph/9906469
\bibitem{gfer96} Giovannini, G. \& Feretti, L. 1996, in {\it
Extragalactic Radio Sources}, R. Ekers et al.  eds, p.333 
\bibitem{jaffe77} Jaffe, W.J. 1977, ApJ, 212, 1
\bibitem{jaffe80} Jaffe, W.J. 1980, ApJ, 241, 924
\bibitem{longair} Longair, M. 1993, `High Energy Astrophysics'
\bibitem{kaa99} Kaastra, J. et al. 1999, ApJ, in press
\bibitem{pxf99} Ping,X.W. et al. 1999, preprint astro-ph/9905106 (PXF99)
\bibitem{re79} Rephaeli, Y. 1979, ApJ, 227, 364
\bibitem{rol81} Roland, J. et al. 1981, A\&A, 100, 7
\bibitem{sar88} Sarazin, C., 1988, 'X-ray emission from clusters of galaxies',
Cambridge Univ. Press
\bibitem{schli87} Schlikaiser,R. et al. 1987, A\&A, 227, 236
\bibitem{sreek} Sreekumar, P. et al. 1996, ApJ, 464, 628 
\bibitem{ve82} Vestrand, W.T. 1982, ICRC Proceedings, OG 3.1-5,97 
\end{thebibliography}
\end{document}